\documentclass[twocolumn,showpacs,preprintnumbers,amsmath,amssymb]{revtex4}

\usepackage{graphicx}
\usepackage{dcolumn}
\usepackage{bm}

\begin{document}

\preprint{APS/123-QED}

\title{Tunable Band Gap in Graphene with a Non-Centrosymmetric Superlattice Potential.}

\author{Rakesh P. Tiwari and D. Stroud}
\affiliation{
Department of Physics, Ohio State University, Columbus, OH 43210\\
}

\date{\today}
\begin{abstract}

We show that, if graphene is subjected to the potential from an external superlattice, a band gap
develops at the Dirac point provided the superlattice potential has broken inversion symmetry.
As numerical example, we calculate the band structure of graphene in the presence of
an external potential due to periodically patterned gates arranged in a triangular or a square graphene superlattice (TGS or SGS) with broken inversion symmetry,
%RPT :: calling it TGS, that is the notation used by cohen et al. changed in the text also
and find that a band gap is created at the original and, in the case of a TGS, the ``second generation'' Dirac point.  This gap, which extends throughout the superlattice Brillouin zone, can be controlled, in principle, by changing the external potential and the lattice constant
of the superlattice.
%RPT :: TS -> TGS
%RAKESH3: I slightly rewrote the abstract to reflect the changes in content.
For a square superlattice of lattice constant 10 nm, we have obtained a gap as large as 65 meV, for gate voltages no larger than 1.5 V.

\end{abstract}

\maketitle

\section{Introduction}

%RAKESH3: I broke the paper into separate sections, since we are no longer constrained by
%the four page limit.

Ever since the synthesis of high-quality graphene\cite{novoselov}, there has been tremendous interest in the properties of this single-layer form of carbon.   Graphene has a honeycomb lattice structure,
with two atoms per primitive cell and a hexagonal Brillouin zone (BZ).  The Fermi energy $E_F$ of homogeneous, neutral graphene lies at the so-called Dirac point, which occurs at high symmetry points
$K$ in the BZ.  In fact, there are two inequivalent Dirac points ${\bf K}_0$ and ${\bf K}_0^\prime$ with
two distinct valleys of excitations.  Near the Dirac point, the density of states is linear
in $|E-E_F|$ and the spectrum of quasiparticle states is well described by the Dirac equation for
massless fermions.  Partly as a result of this electronic structure\cite{novoselov,zhang,berger,geim}, graphene has many unusual
electronic properties, such as a unique type of quantum Hall effect\cite{novoselev1,purewal}, ballistic conduction by massless Dirac fermions\cite{novoselev1,purewal}, size-dependent band gap\cite{han}, large magnetoresistance\cite{hill,geim,cho}, and gate-tuneable optical transitions\cite{wang}.  
  
A number of workers have investigated the possibility
of building graphene electronic circuits without physically cutting or etching the graphene monolayer.
A natural way to accomplish this is by subjecting graphene to an external potential with a suitable superlattice periodicity, e.\ g., by applying appropriate gate voltages.  Such superlattices have proven to be extremely successful in controlling the electronic structure of more 
conventional semiconducting materials (see, e.\ g., Ref.\ \cite{tsu}).  In these materials,
the presence of the additional periodic potential gives rise to superlattice electronic band structure, which has extra band gaps 
at high symmetry points in the superlattice Brillouin zone (SBZ).

There have been several predictions of electronic effects in graphene
due to an external superlattice potential.  For example, with  
one-dimensional (1D) and two-dimensional (2D) superlattices, the group velocity of the low-energy charge carriers is anisotropically renormalized \cite{ar}, while a corrugated 
graphene sheet is expected to show charge inhomogeneity and localized states \cite{guinea}. 
Superlattices in graphene can be realized experimentally by using periodically patterned gates. 
Using an electron-beam, adatoms with a superlattice patterns of periodicity 
as small as 5 nm have been achieved on freestanding graphene membranes 
\cite{meyer}. Superlattice patterns have also been observed for graphene on metal surfaces \cite{marchini,vazquez,pan}. 
%RAKESH, the above two are experimental results, correct?  What do you mean by ``superlattice
%patterns have been observed''?
%comment:: yes they are experimental papers the way you have phrased it is better
In a recent theoretical study, a triangular graphene superlattice (TGS) was considered, and a new class of massless Dirac fermions was predicted to 
occur at the M point in the SBZ\cite{park}.

In the studies so far, the external periodic potential giving rise to the superlattice 
%RAKESH2: removed the word ``all'' before studies in the line above (out of caution)
has had spatial inversion symmetry.
%RPT:: removed on extra "has"
As a result of this symmetry, the degeneracy of the conduction and valence band at the original Dirac point (ODP) 
is preserved. In the present work, by contrast, we consider the potentials of an external TGS and a square graphene
superlattice (SGS) {\it without} inversion symmetry.  We show 
that, because of the absence of inversion symmetry, an energy gap opens up at both the original and the new Dirac points in the TGS and at the Dirac point in the SGS.  The magnitude of 
these gaps can, in principle, be controlled by modifying the externally applied voltages.  Thus far, we have been
able to achieve a band gap as large as 65 meV extending throughout the SBZ.
%RAKESH3: please fill in the value here. this value. 
Because of the controllability of the band gap, this type of
graphene superlattice represents a system in which the band gap 
can be chosen with considerable freedom by modifying the applied external potential.  Such a system, and particularly the control at the ODP,  might be quite useful in realizing future graphene electronic circuits.

\section{Formalism}

We first describe the formalism we use to calculate the superlattice band structure for a non-centrosymmetric superlattice potential, mostly following the approach of Ref.\ \cite{park}.
If the periodicity of the superlattice is much larger than the intercarbon distance 
$a \sim$ 1.42 $\AA$, which is the case considered here, the intervalley scattering 
(between {\bf K}$_0$ and {\bf K}$^{\prime}_0$) can be neglected \cite{ando}. We limit our discussion to the spectrum near one of the two inequivalent Dirac points, which we denote
${\bf K}_0$, in the presence of a periodic external potential $V(x,y)$.  In pure graphene, we use a pseudospin basis $\left( \begin{array}{cc}1 \\ 0 \end{array} \right)e^{i{\bf k}\cdot{\bf r}}$ and
$\left(\begin{array}{cc}0\\1\end{array}\right)e^{i{\bf k}\cdot{\bf r}}$, where 
$\left( \begin{array}{cc} 1\\0 \end{array} \right)$ and 
$\left(\begin{array}{cc}0\\1\end{array}\right)$ are Bloch sums of $\pi$ orbitals with wave vector
${\bf K}_0$ on the sublattices A and B, respectively, and ${\bf k}$ is the wave vector relative to
the ${\bf K}_0$ point.  The single valley Hamiltonian for the quasiparticles of pure graphene is\cite{wallace}
\begin{equation}
H_0 = \hbar v_0(-i\sigma_x\partial_x - i\sigma_y\partial_y),
\end{equation}
where $v_0$ is the (isotropic) group velocity at the Dirac point and the $\sigma$'s are Pauli matrices.  The eigenstates and energy eigenvalues are
\begin{equation}
\psi_{s,{\bf k}}^0({\bf r}) = 
\frac{1}{\sqrt{2}}\left(\begin{array}{cc}1\\se^{i\theta_{\bf k}}\end{array}\right)
\label{eq:wf}
\end{equation}
and
\begin{equation}
E_s^0({\bf k}) = s\hbar v_0k,
\end{equation}
where $s = \pm 1$ is the band index and $\theta_{\bf k}$ is the polar angle of the wavevector ${\bf k}$.  Eq.\ (\ref{eq:wf}) indicates that the pseudospin vector is parallel 
and antiparallel to ${\bf k}$ in the upper ($s=1$) and lower ($s=-1$) bands, respectively.  

We wish to consider the spectrum of elementary excitations near one of the Dirac points in the
presence of $V(x, y)$.  The total Hamiltonian then takes the form
$H = H_0 + H^\prime$, where
\begin{equation}
H^\prime = IV(x,y),
\end{equation}
$I$ being the $2 \times 2$ identity matrix.  We consider a periodic external potential having either a triangular Bravais lattice with basis vectors ${\bf b}_1 = b\hat{x}$, ${\bf b}_2 = b\left(\frac{1}{2}\hat{x} + \frac{\sqrt{3}}{2}\hat{y}\right)$, or a square Bravais lattice with ${\bf b}_1 = b\hat{x}$,
${\bf b}_2 =b\hat{y}$.  
%RAKESH3: I added a half-sentence about the square lattice also.
%RAKESH: I thought better to use b for superlattice constant, a for graphene lattice
%constant.  Please check to make sure I have done this consistently. 
%comment:: I have changed the figures and it is now consistent.
We also assume that $V(x,y)$ varies slowly on the scale of a lattice 
constant ($b \gg a$).  
Then the band structure of the elementary excitations near the {\bf K}$_0$ point can readily be obtained by diagonalizing $H_0 + H^\prime$ in a plane
wave spinor basis\cite{park}.   The basis states
%RAKESH3: corrected the phrase ``The members of the basis state''
are of the form $\chi(s)\exp[i({\bf k} + {\bf G})\cdot {\bf r}]$, where 
$\chi(s) = \frac{1}{\sqrt{2}}\left( \begin{array}{cc} 1\\e^{i\theta_{{\bf k} + {\bf G}}}\end{array}\right)$ or 
$\frac{1}{\sqrt{2}}\left(\begin{array}{cc}1\\-e^{i\theta_{{\bf k}+{\bf G}}}\end{array}\right)$ and ${\bf G}$ is a reciprocal lattice vector corresponding to the periodic external potential $V(x, y)$.  If we include $N$ plane waves, we must
thus diagonalize a $2N \times 2N$ matrix to obtain the band structure of elementary excitations 
non-perturbatively
near the ${\bf K}_0$ point in the presence of the external periodic potential.  We use $N = 625$ for both the
triangular and the square graphene superlattice.
%RPT :: K_0 is bold
%comment:: we are calculating the band structure non perturbatively, so i changed few words
%RAKESH3: Did you use N = 625 for the square lattice too?
%NEW: YES
In our actual calculations, we have considered an external potential produced by a periodic array of circular regions, within each of which the potential is a constant. 
These are easy to treat by the above approach, because the Fourier transform of the constant potential is available analytically.  To break the inversion symmetry, we have added a second, smaller circular region within each primitive cell,
as illustrated in Figs.\ 1 and 2.  If the second region is not equidistant from
the two adjacent larger circular regions, inversion symmetry is broken.

\section{Results}

We first review the case of a triangular
array of large circles, with a constant external potential within each circle.
Elsewhere in the superlattice, the external potential is zero.  The resulting band structure
has been computed in Ref.\ \cite{park}.
In the presence of the superlattice, there are several Dirac points: the ODP, which is at $\Gamma$ of 
the SBZ, and ``next generation'' Dirac points, at $M$ and $K$ of the SBZ.  Of
these new Dirac points, only the ODP (with energy set to 0 eV) and the new Dirac point (NDP) at M with energy 0.196 eV) have vanishing 
density of states\cite{park}.  Near both of these points, the density of states goes linearly to zero 
with energy. However, the group velocity near the NDP is highly anisotropic.

We now consider the superlattice band structure (SBS) for graphene subject to
the non-centrosymmetric potential shown in Fig.\ 1.   We choose the lattice constant of the
external potential to be $b=10$ nm, and the radius $r_1$ of the large circles to be $2.5$ nm, so that the filling fraction of large circles is 0.226725.  We take the smaller circular gates to have radius $r_2=1$ nm, and to be centered at ${\bf R} + 0.4b{\bf \hat{x}}$, where ${\bf R}$ is a Bravais
vector of the superlattice.  The resulting band structure is shown in Fig.\ 3 for the case 
$V_1 = 0.5$V, $V_2 = -1.65$V, where $V_1$ and $V_2$ are the voltages on the larger and smaller circles.
%RAKESH3: what does V = 0 represent?  I suppose it is the voltage on the rest of the graphene.
%RAKESH3: Did you change the symbols for the voltages to $V_1$ and $V_2$ rather than $V_0$ and $V_1?
%NEW:: YES 
Once again $V = 0$ in the rest of the lattice.
A substantial direct gap opens up at the ODP of magnitude $\Delta_\Gamma = 58.2$ meV for these parameters (between the bands above and below the ODP).  
Although not obvious from Fig.\ 3, there is also a gap $\Delta_M$ which opens up at the NDP, of magnitude $\sim 1$ meV.
%RAKESH3: I added the above lines.  Note the phrase ``direct band gap''. 
For these parameters, we have checked that there is, in fact, a full band gap at ODP extending through the entire SBZ. 
%NEW:: I am commenting these lines out as the gap is direct
%The minimum band gap for these parameters is indirect.  The maximum of the ``valence
%band'' of the SBS occurs at $\Gamma$ and the bottom of the ``conduction band'' lies along the line $\Gamma$....;
%RAKESH3: please fill in here.  Is it along the line Gamma-M?
%its magnitude is ....
%RAKESH3: please fill in the number here.
%The valence and conduction bands are taken to be those bands which meet at the Dirac point in the absence of a
%non-centrosymmetric perturbation.
The small gap also at $M$ does not extend throughout the SBZ, as is evident from the Figure (the gap at $M$ does extend throughout the SBZ for $V_2$ positive).  
The inset of Fig.\ 3 shows the SBZ, with the symmetry points $\Gamma$, $K$, and $M$ indicated.  

%%%%%%%%%%%%%%%%%%%%%%%%%%%%%%%%%%%%%%%%%%%%%%%%%%%%%%%%%%%%%%%%%%%%%%%%%%%%%%%%%%%%%%%%%%%%%%%%%%%%%%%%%
%%%%%%%%%%%%%%%%%% This is added by Rakesh P Tiwari on Apr 16 %%%%%%%%%%%%%%%%%%%%%%%%%%%%%%%%%%%%%%%%%%
%%%%%%%%%%%%%%%%%%%%%%%%%%%%%%%%%%%%%%%%%%%%%%%%%%%%%%%%%%%%%%%%%%%%%%%%%%%%%%%%%%%%%%%%%%%%%%%%%%%%%%%%

In Fig.\ 4, we show the corresponding SBS for graphene subjected to the periodic 
potential with square symmetry shown in Fig.\ 2.  
Once again, we choose the superlattice constant $b = 10$ nm, $r_1 = 2.5$ nm,
and $r_2 = 1$ nm, so that the filling fraction of circles in this case is 0.227765.  We
choose $V_1 = 0.5$ V and $V_2 = -1.4$ V, and each small circle is located at $0.375\hat{x}$ relative
to the center of each large circle.   There is a gap of magnitude $64.4$ meV, at the ODP at $\Gamma$
which extends throughout the SBZ. 
%The minimum gap is again indirect; the maximum of the valence band
%occurs at $\Gamma$ and the conduction band minimum along the line 
%the line $\Gamma$....
%RAKESH3: please fill in here.

%%%%%%%%%%%%%%%%%%%%%%%%%%%%%%%%%%%%%%%%%%%%%%%%%%%%%%%%%%%%%%%%%%%%%%%%%%%%%%%%%%%%%%%%%%%%%%%%%%%%%%%%
%%%%%%%%%%%%%%%%%%%%%%%%%%%%%%%%%%%%%%%%%%%%%%%%%%%%%%%%%%%%%%%%%%%%%%%%%%%%%%%%%%%%%%%%%%%%%%%%%%%%%%%

In Fig.\ 5, we show show how the direct gaps at the Dirac point depend on the voltage $V_2$ on the smaller
circles in the triangular and square superlattices.  For the square superlattice, 
we take $b = 10$ nm, $r_1/b = 0.25$ $r_2/b = 0.125$, $V_1$ at $0.5$ V, and the small circles
are located at $0.375\hat{x}$.  For the
triangular superlattice, $b = 10$ nm, $r_1/b = 0.25$, $r_2/b = 0.1$, $V_1 = 0.5$ V, and the small circles are located at $0.4\hat{x}$.  In both cases, these gaps show clear maxima as functions of $V_2$.  We have carried out
similar calculations using positive values of $V_2$ in both cases and also obtain
nonzero gaps extending throughout the SBZ.  In the case of positive $V_2$, they are generally smaller than for negative voltages.
%RAKESH3: Is the above statement correct?  I just guessed.
%NEW: yes

In Fig.\ 6, we show the dependence of the gaps at the Dirac point on the
radius $r_2$ in both the triangular and square superlattices.  For the square
superlattice, we take $b = 10$ nm, $V_1 = 0.5$ V, $V_2 = -1.4$ V, $r_1/b = 0.25$ and the small
circles are located at $0.375\hat{x}$.  For the triangular case, $b = 10$ nm, $V_1 = 0.5$ V,
$V_2 = -1.65$ V, $r_1/b = 0.25$, and we locate the small circular contacts at $0.4\hat{x}$.
The direct band gaps can reach as large as  $\sim$ 65 meV for the square superlattice.  
%However,
%the full band gap, which is indirect as noted above, is closer to 40 meV for this
%choice of lattice constant.
%RAKESH3, have I said this right?  Are the facts correct?
%NEW: yes

Because of the special form of the Hamiltonian of eqs.\ (1) and (4), the superlattice band
structure, and hence the band gap at the ODP, exhibits a simple scaling relation: under
the transformation $b \rightarrow \lambda b$ and $V(x, y) \rightarrow V(\lambda x, \lambda y)/\lambda$, where $\lambda$ is a dimensionless scaling 
factor, the eigenvalues of $H$ satisfy
$E(k_x, k_y) \rightarrow E(k_x/\lambda, k_y/\lambda)/\lambda$.  Thus, in particular, $\Delta_\Gamma$ is multiplied by $1/\lambda$. 
For example, if $b$ is halved (and all other lengths are also halved), and the potentials on all the contacts are doubled, then the band gap is also 
doubled.  We have confirmed numerically that this scaling relation is satisfied for our bands.
 
%RAKESH2: above paragraph slightly reworded.

%We have also confirmed another relation numerically, namely, that for fixed $V_1$, $r_1$, and $r_2$, and for small $V_2$, $\Delta_\Gamma \propto 
%|V_2|$.  This relation can be derived analytically, for small $V_1$, by applying first-order perturbation theory to the two degenerate states
%at the superlattice point $\Gamma$, using as a perturbation the potential produced by the small
%off-center circular contacts. 
%RAKESH3: I rewrote and shortened this paragraph, since we no longer have the old figures with the
%contour plots.

Besides these calculations, we have tried some other geometries in an effort to obtain the largest possible band gap. For example, we have calculated the SBS for triangular and square superlattices with two off-center circular
contacts, one along each edge of the primitive cell. This geometry does not lead to band gaps larger than we have obtained with one off-center circle.  
%RAKESH3: any other geometries I should mention?

\section{Discussion}

The present results show that the band gap of graphene at the Dirac point can be manipulated by subjecting it to a superlattice potential which lacks inversion symmetry. 
The magnitude of the band gap can be controlled by modifying the contact potentials.  Thus, if
such an arrangement of contacts can be created, the resulting material has a band gap which can be
controlled via the applied voltages without modifying the structure.

There are several obvious challenges before this scheme could be used in practice.
First, the band gaps are rather small (of order 65 meV).  
%RAKESH - what is the largest band gap you can reach?
These can be made larger by reducing
$b$, while simultaneously increasing $V_1$ and $V_2$  
as implied by the above scaling relation,
but this could be challenging experimentally.
The gaps can also be increased by increasing the voltages at fixed geometry.   Perhaps the most promising scheme might be to increase $b$ while simultaneously {\it increasing} the bias voltages.  The larger $b$ would be easier to achieve experimentally, while the large voltage offsets would increase the gap.  These ideas certainly do not exhaust the possibilities offered by periodic arrangements of contacts.  Any 2D superlattice of gates lacking inversion symmetry would lead to a
band gap at the ODP, and some which we have not tried may lead to a larger band gap than those we have found to date.  A 1D superlattice potential lacking inversion symmetry will not suffice to produce a complete band gap.  But such a 1D potential, when combined with a suitable non-time-reversal-invariant perturbation such as a magnetic field, might also lead to a complete band gap.
%RAKESH, do you agree with the above statement?  It seems to me one needs a 2D potential.

There are, of course, many other ways to create a band gap in graphene besides
the method described here.  For example, one could create a non-centrosymmetric
lattice of nanoscale holes in graphene.  It would be difficult to use this method, however, to create a tunable band gap.  In graphene nanoribbons with armchair edges, the band structure becomes insulating if the width of the 
sample, in units of the lattice constant $a$, is not of the form $3M+1$, with $M$ an 
integer\cite{breyfertig,son,ezawa,barone,chen}.
%RAKESH3: I added the references suggested by the referee   
But creating such a ribbon requires cutting graphene samples with very high precision, and again,
once the ribbon has been created, the band gap cannot be easily tuned.  Other possibilities are to induce a gap by using a substrate which lacks 
inversion symmetry\cite{giovannetti}, or to use graphene antidot lattices\cite{pedersen} Again in this case, the gap cannot easily be controlled because the atomic spacing of the substrate is fixed.  Our proposal of a TGS
(or other 2D superlattice) with a non-centrosymmetric superlattice potential is more efficient in opening up a gap at the ODP.
In principle, moreover, this gap can easily be larger than the thermal 
energy at room temperature for modest values of potential.   Thus, this method may be a viable approach to creating graphene with a readily tunable band gap.
%RAKESH1: mention the case of graphene on a SiC or BN substrate?  I read about this somewhere but
%can't find it now.

In summary, we have calculated the superlattice band structure for graphene subjected to a non-centrosymmetric
superlattice of contacts, on each of which the voltage is held constant.  We find that the superlattice band structure
exhibits a band gap extending throughout the SBZ, which can thus be tuned by external voltages.  
For some choices of the voltages and superlattice constant, the band gap can be as large as $65$ meV, 
significantly larger than the room temperature thermal energy.  
Thus, this arrangement might possibly be of use in future electronic or electromagnetic devices.

\section{Acknowledgments}

This work has been supported in part by NSF DMR-0820414 through the Materials Research Science and Engineering Center at Ohio State University.  We acknowledge valuable conversations with J.\ P.\ Pelz.
%RAKESH2: should be acknowledge Jon Pelz?  Anyone else?

\newpage

%RAKESH2: in the figure below, the notation is inconsistent with that used in the text.
%For example, I believe that R should be r_1, r should be r_2, a should be b, and b should
%be 0.4b.  Or perhaps I don't have the latest version of the figure?
\begin{figure}[h]
\begin{center}
\includegraphics[scale=0.4,angle=0]{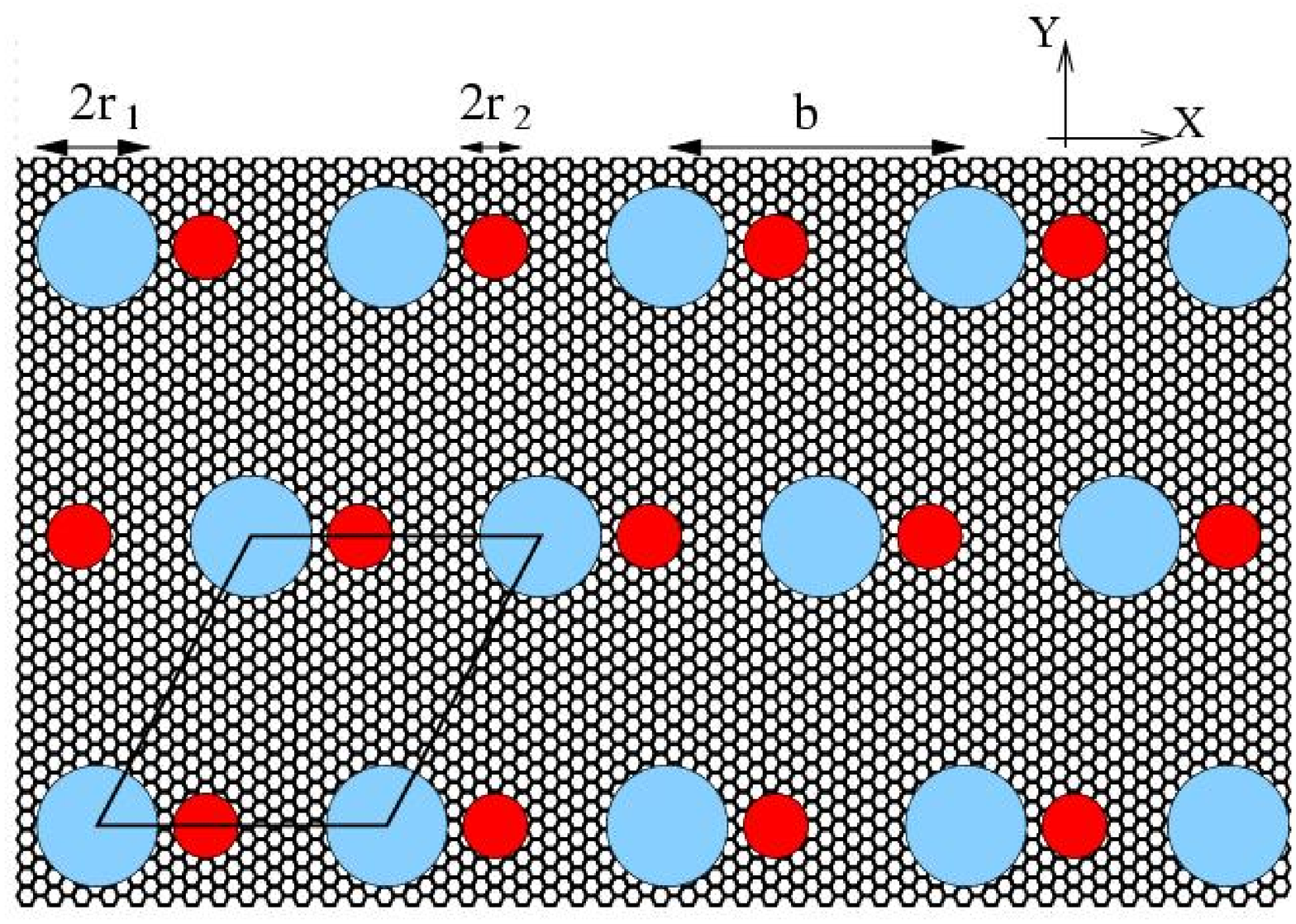}
\caption{(Color online.)  Schematic of graphene plane subjected to a periodic external potential $V(x, y)$ consisting of a periodic arrangement 
of circular contacts, within each of which $V$ is a constant.
The large circles are arranged on a triangular lattice of edge $b$.  The small circles are not midway 
between the large circles; for the case shown, they are centered at $0.4b\hat{x}$
relative to the centers of the large circles.}
%RAKESH: I am using b, not a, for the lattice constant of the superlattice.  Also, the distance between
%the centers of the small and the large circles should be something other than b.  How about 0.4, which is
%the value used in the text?
%comment:: i have changed the figure and removed the distance between bigger and smaller circles we can just mention it is not midway
% I have added coordinate axes and changed caption
\end{center}
\label{fig:1}
\end{figure}

\begin{figure}[h]
\begin{center}
\includegraphics[scale=0.4,angle=0]{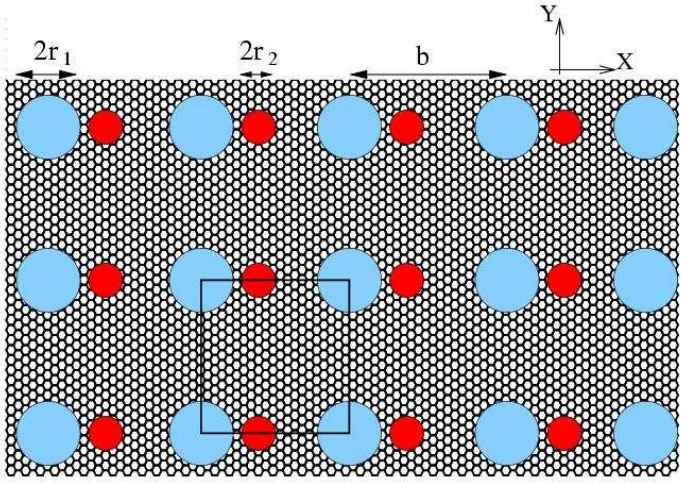}
\caption{(Color online.)  Same as Fig.\ 1, except that the large and small circular contacts
are arranged on a square lattice.  Each small circle is now centered at $0.375b\hat{x}$ relative to the center of the nearest large circle.}
\end{center}
\label{fig:2}
\end{figure}

\begin{figure}[t]
\begin{center}
\includegraphics[scale=0.5,angle=270]{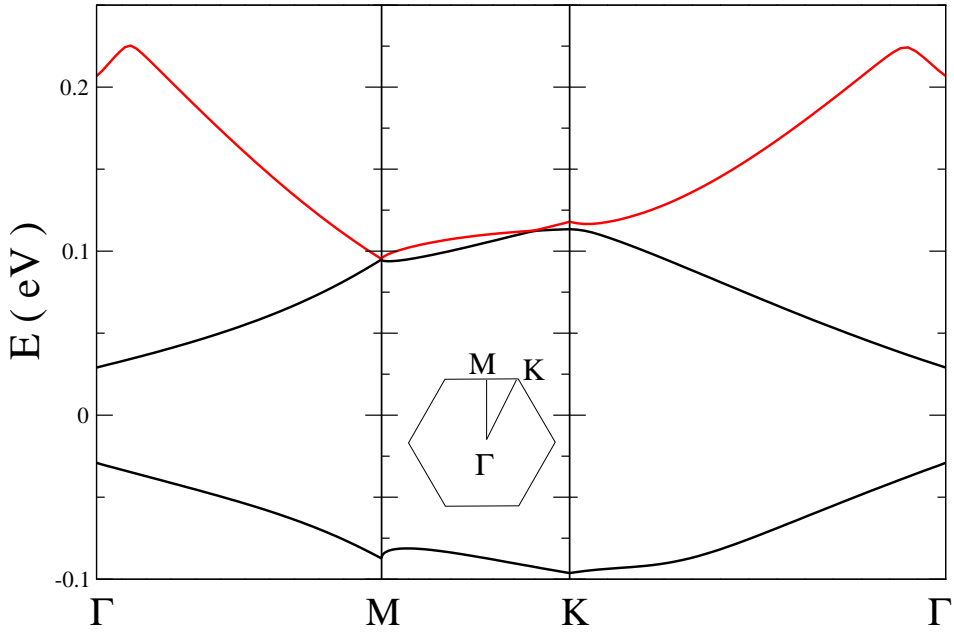}
\caption{(Color online) Superlattice band structure (SBS) for graphene subjected to an external potential produced by a triangular superlattice of contacts, as shown in Fig.\ 1.  The calculation is
carried out using $b=10$ nm, $V_1=0.5$ V, $V_2=-1.65$ V, $r_1/b=0.25$, $r_2/b=0.1$, and the small circles are located at $0.4\hat{x}$.   The $\Gamma$ point of the superlattice corresponds to one of the K point of the original graphene lattice.  
If only the large circular contacts were present, the SBS
would have Dirac points at both $\Gamma$ and $M$ of the SBS Brillouin zone, at both of which
the density of states would go linearly to zero.  
%If both large and small
%circular contacts are present, inversion symmetry is broken.  A small gap opens up at both the %original Dirac point at $\Gamma$ of the superlattice band 
%structure and the superlattice Dirac points at $M$.  However, for these voltages, the SBS no %longer has a vanishing density of states, near $M$, as is evident from the 
%Figure.  
%RAKESH3: removed above six lines from caption, as already stated in the text.
Inset: first Brillouin zone of a triangular lattice showing the three high-symmetry points $\Gamma$, $M$, and $K$.}
\end{center}
\label{fig:3}
\end{figure}

\begin{figure}[t]
\begin{center}
\includegraphics[scale=0.5,angle=270]{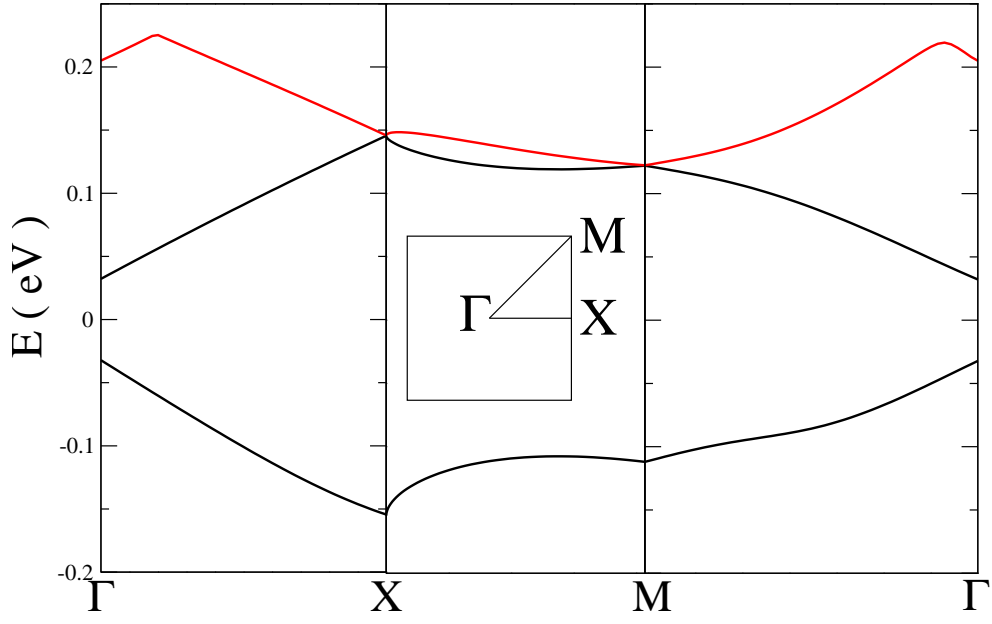}
\caption{(Color online) SBS for graphene subjected to an external potential
produced by a square superlattice of contacts, as in Fig.\ 2.  The $\Gamma$ point of the superlattice corresponds to one of the K point of the 
original graphene lattice.  
%If only the large circular contacts were present, the SBS
%would have Dirac points at $\Gamma$ of the SBS, at which
%the density of states would go linearly to zero.  If both large and small
%circular contacts are present, inversion symmetry is broken, and a small gap opens up at the %Dirac point at $\Gamma$ of the
%superlattice Brillouin zone, as shown in the
%Figure. 
%RAKESH3: removed above seven lines from caption. 
In this case, we use $b=10$ nm, $V_1 =0.5$ V, $V_2=-1.4$ V, $r_1/b=0.25$, $r_2/b=0.125$ and the small circles are located at $0.375\hat{x}$.  Inset: first 
Brillouin zone of a square lattice showing the three high-symmetry points $\Gamma$, $X$, and $M$.}
\end{center}
\label{fig:4}
\end{figure}

\begin{figure}[t]
\begin{center}
\includegraphics[scale=0.5,angle=270]{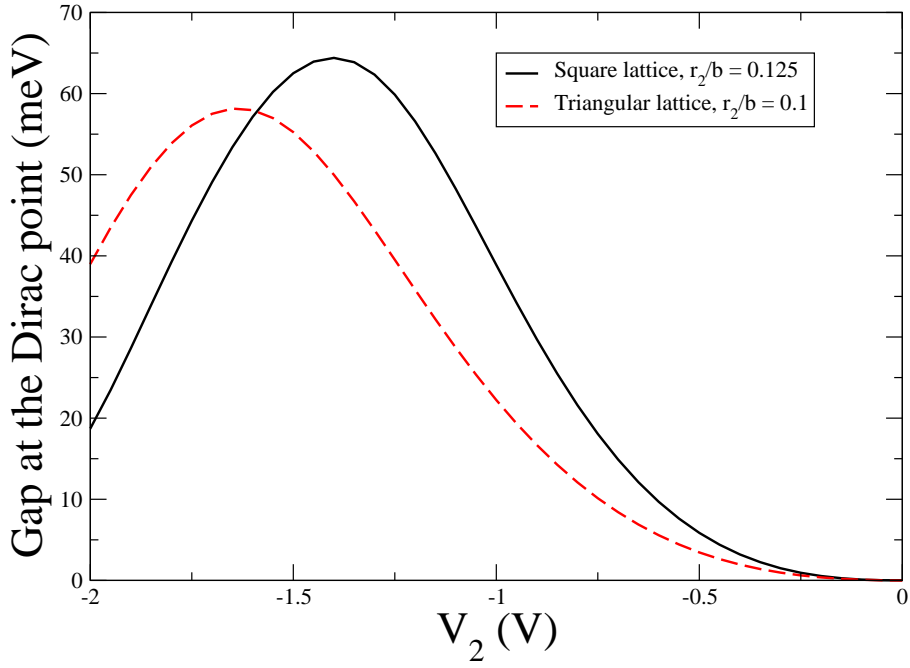}
\caption{(Color online) Plot of the gap at the Dirac point $\Gamma$
%RAKESH3: inserted ``\Gamma'' in line above
as a function of the voltage $V_2$ (in V) on the smaller circles.  Full curve: square lattice; dashed curve: triangular lattice. For the square lattice, we use $b=10$ nm, $V_1 =0.5$ V, $r_1/b=0.25$, $r_2/b=0.125$ and 
small circles are located at $0.375\hat{x}$.  For the triangular lattice, $b=10$ nm, $V_1=0.5$ V, $r_1/b=0.25$, $r_2/b=0.1$ and the small 
circles are located at $0.4\hat{x}$.}
\end{center}
\label{fig:5}
\end{figure}
\begin{figure}[t]
\begin{center}
\includegraphics[scale=0.5,angle=270]{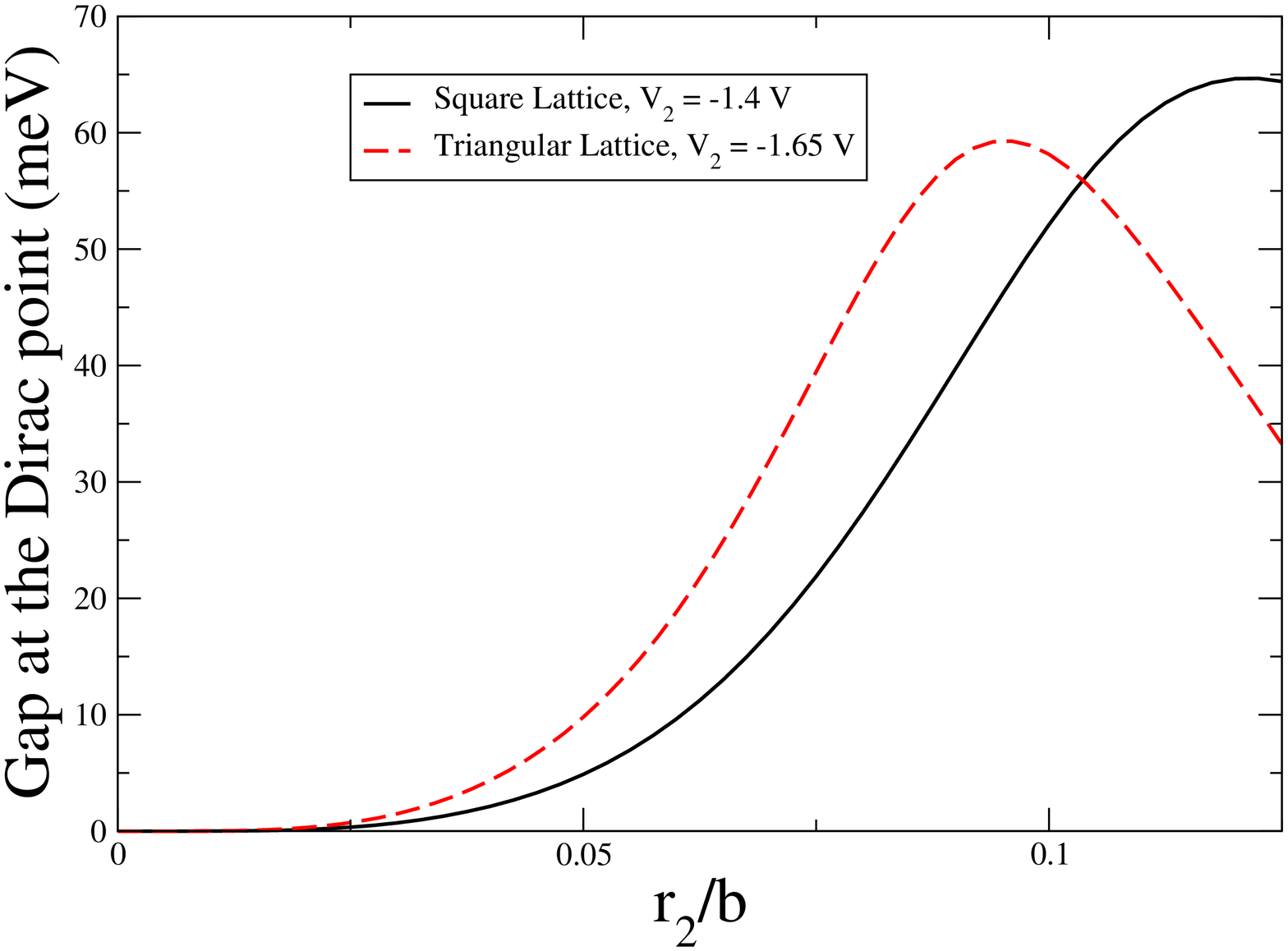}
\caption{(Color Online) Plot of the gap at the Dirac point as a function $r_2/b$.
Full curve: square lattice; dashed curve: triangular lattice. For the square lattice $b=10$ nm, $V_1 =0.5$ V, $V_2=-1.4$ V, $r_1/b=0.25$, and
small circles are located at $0.375\hat{x}$.  For the triangular lattice, $b=10$ nm, $V_1=0.5$ V, $V_2=-1.65$ V, $r_1/b=0.25$ and small
circles are located at $0.4\hat{x}$.}
\end{center}
\label{fig:6}
\end{figure}


\newpage



\begin{thebibliography}{99}

\bibitem{novoselov} K. S. Novoselov, A. K. Geim, S. V. Morozov, D. Jiang, M. I. Katsnelson, I. V. Grigorieva, S. V. Dubonos and A. A. Firsov, Nature (London) {\bf 438}, 
197 (2005).

\bibitem{zhang} Y.\ Zhang, J.\ W.\ Tan, H.\ L.\ Stormer and P.\ Kim, Nature (London) {\bf 438}, 201 (2005).

\bibitem{berger} C.\ Berger, Z.\ Song, X.\ Li, X.\ Wu, N.\ Brown, C.\ Naud, D.\ Mayou, T.\ Li, J.\ Hass, A.\ N.\ Marchenkov, E.\ H.\ Conrad, P.\ N.\ First, W.\ A.\ de Heer, 
Science {\bf 312}, 1191 (2006).

\bibitem{geim} A.\ K.\ Geim and K.\ S.\ Novoselov, Nature Mater. {\bf 6}, 183 (2007).

\bibitem{novoselev1} V.\ S.\ Novoselev, A.\ K.\ Geim, S.\ V.\ Morozov, D.\ Jiang, M.\ I.\ Katsnelson, I.\ V.\ Grigorieva, S.\ V.\ Dubonos and A.\ A.\ Firsov, Nature {\bf 438}, 197 (2005). 

\bibitem{purewal} M.\ S.\ Purewal, Y.\ Zhang, and P.\ Kim, Phys.\ Stat.\ Solidi B {\bf 243}, 4318 (2006).

\bibitem{han} M.\ Y.\ Han, B.\ Ozyilmaz, Y.\ Zhang, and P.\ Kim, Phys.\ Rev.\ Lett.\ {\bf 98}, 206805 (2007).

\bibitem{hill} E.\ W.\ Hill, A.\ K.\ Geim, K.\ S.\ Novoselev, F.\ Schedin,and P.\ Blake, IEEE Trans.\ Magn. {\bf 42}, 2694 (2008).

\bibitem{cho} S.\ Cho and M.\ S.\ Fuhrer, Phys.\ Rev.\ {\bf B} {\bf 77}, 081402(R) (2008).

\bibitem{wang}  F.\ Wang, Y.\ Zhang, C.\ Tian, C. Girit, A.\ Zettl, M.\ Crommie, and Y.\ R.\ Shen, Science {\bf 32]}, 206 (2008).

\bibitem{tsu} R.\ Tsu, \textit{Superlattice to Nanoelectronics} ( Elsevier, Oxford, 2005).

\bibitem{ar} C.-H.\ Park \textit{et al.}, Nature Phys. {\bf 4}, 213 (2008); M.\ Barbier, F.\ M.\ Peeters, P.\ Vasilopoulos and J.\ M.\ Pereira, Jr., Phys. Rev. B {\bf 77}, 
115446 (2008).

\bibitem{guinea} F.\ Guinea, M.\ I.\ Katsnelson and M.\ A.\ H.\ Vozmediano, Phys. Rev. B {\bf 77}, 075422 (2008).

\bibitem{meyer} J.\ C.\ Meyer, C.\ O.\ Girit, M.\ F.\ Crommie and A.\ Zettl, Appl. Phys. Lett. {\bf 92}, 123110 (2008).

\bibitem{marchini} S.\ Marchini, S.\ G\"unther and J. Wintterlin, Phys. Rev. B {\bf 76}, 075429 (2007).

\bibitem{vazquez} A.\ L.\ Vazquez de Parga, F.\ Calleja, B.\ Borca, M.\ C.\ G.\ Passeggi, Jr., J.\ J.\ Hinarejos, F.\ Guinea and R.\ Miranda, Phys. Rev. Lett. {\bf 100}, 
056807.

\bibitem{pan} Y. Pan \textit{et al.}, arXiv:0709.2858v1.
%RAKESH2: Has this been published?  This is more than a year old.


\bibitem{park} C.-H.\ Park, L.\ Yang, Y.-W.\ Son, M.\ L.\ Cohen, and S.\ G.\ Louie, Phys.\ Rev.\
Lett.\ {\bf 101}, 126804 (2008)

\bibitem{ando} T.\ Ando and T.\ Nakanishi, J.\ Phys. Soc. Jpn. {\bf 67}, 1704 (1998).

\bibitem{wallace} P.\ R.\ Wallace, Phys.\ Rev. {\bf 71}, 622 (1947).

\bibitem{breyfertig} L.\ Brey and H.\ A.\ Fertig, Phys. Rev. B {\bf 73}, 235411 (2006).

\bibitem{son} Y.-W.\ Son, M.\ L.\ Cohen, and S.\ G.\ Louie, Phys.\ Rev.\ Lett.\ {\bf 97}, 216803 (2006).

\bibitem{ezawa} M.\ Ezawa, Phys.\ Rev.\ {\bf B} {\bf 73}, 045432 (2006).

\bibitem{barone} V.\ Barone, O.\ Hod, and G.\ E.\ Scuseria, Nano Letters {\bf 6}, 2748 (2006).

\bibitem{chen} Z.\ Chen, Y.-M.\ Lin, M.\ J.\ Rooks, and P.\ Avouris, Physica E {\bf 40}, 228 (2007).

\bibitem{giovannetti} G.\ Giovannetti, P.\ A.\ Khomyakov, G.\ Brocks, P.\ J.\ Kelly,
and J.\ van den Brink, Phys.\ Rev.\ {\bf B} {\bf 76}, 073103 (2007).

\bibitem{pedersen} T.\ G.\ Pedersen, C.\ Flindt, J.\ Pedersen, N.\ A.\ Mortensen, A.-P.\ Jauho, and K.\ Pedersen,
Phys.\ Rev.\ Lett.\ {\bf 100}, 136804 (2008).

\end{thebibliography}
\end{document}